\documentclass[12pt]{article}
\usepackage{amssymb, amsthm, amsmath, natbib, bm, color, xcolor, graphicx, epstopdf, soul}
\newcommand{\Nystrom}{Nystr{\"o}m }

\newcommand{\bvS}{\bm{\varSigma}}
\newcommand{\hbvS}{\widetilde{\bm{\varSigma}}}
\newcommand{\btt}{\bm{\theta}}
\newcommand{\mle}{\widehat{\bm{\theta}}}

\newcommand{\eps}{\varepsilon}            
\newcommand{\ccg}[1]{{\color{blue}}} 

\renewcommand\hat{\widehat}
\renewcommand{\baselinestretch}{1.2}
\newcommand\bx{\bm{x}}
\newcommand\by{\bm{y}}

\newcommand{\revs}[1]{{\textcolor{black}{#1}}}

\begin{document}

\def\spacingset#1{\renewcommand{\baselinestretch}%
{#1}\small\normalsize} \spacingset{1}

\font\afffont=cmr12 at 12pt 

\title{\bf Scalable Gaussian Process Computations Using Hierarchical Matrices}

\author{\textbf{Christopher J. Geoga} \thanks{corresponding author: \texttt{cgeoga@anl.gov}} \\
        \afffont Mathematics and Computer Science Division, Argonne National Laboratory\\
        \and
        \textbf{Mihai Anitescu}
        \thanks{This material was based upon work supported by the U.S.  Department of Energy,
          Office of Science, Office of Advanced Scientific Computing Research (ASCR) under Contract
          DE-AC02-06CH11347. We acknowledge partial NSF funding through awards FP061151-01-PR and
          CNS-1545046 to MA.} \\
        \afffont Mathematics and Computer Science Division, Argonne National Laboratory\\
        \afffont Department of Statistics, University of Chicago \\
        \and
        \textbf{Michael L. Stein}
        \thanks{This material was based upon work supported by the U.S.  Department of Energy,
          Office of Science, Office of Advanced Scientific Computing Research (ASCR) under Contract
          DE-AC02-06CH11357.} \\
        \afffont Department of Statistics, University of Chicago 
      }

\date{}

\maketitle

\bigskip

\begin{abstract}
We present a kernel-independent method that applies hierarchical matrices to the problem of maximum
likelihood estimation for Gaussian processes.  The proposed approximation provides natural and
scalable stochastic estimators for its gradient and Hessian, as well as the expected Fisher
information matrix, that are computable in quasilinear $O(n \log^2 n)$ complexity for a large range
of models. To accomplish this, we (i) choose a specific hierarchical approximation for covariance
matrices that enables the computation of their exact derivatives and (ii) use a stabilized form of
the Hutchinson stochastic trace estimator.  Since both the observed and expected information
matrices can be computed in quasilinear complexity, covariance matrices for MLEs can also be
estimated efficiently.  \revs{In this study, we demonstrate the scalability of the method, show how
details of its implementation effect numerical accuracy and computational effort}, and validate that
the resulting MLEs and confidence intervals based on the inverse Fisher information matrix
faithfully approach those obtained by the exact likelihood. 
\end{abstract}


\newpage


\section{Introduction}

Many real-valued stochastic processes $Z(\bm{x})$, $\bx \in \mathbb{R}^d$,  are modeled as Gaussian
processes, so that observing $Z(\bm{x})$ at locations $\{ \bm{x}_{j} \}_{j=1}^{n}$ results in the
data $\bm{y} \in \mathbb{R}^n$ following a $N(\bm{\mu}, \bvS)$ distribution.  Here the covariance
matrix $\bvS$ is parameterized by a valid covariance function $K(\cdot,\cdot;\btt)$ that depends on
parameters $\btt \in \mathbb{R}^m$, so that
\begin{align*}
  \bvS_{j,k} & = \text{Cov}(Z(\bm{x}_j), Z(\bm{x}_k)) \\
             & = K(\bm{x}_j, \bm{x}_k; \btt), \quad j,k=1,2,\ldots,n.
\end{align*}

In many cases in the physical sciences, estimating the parameters $\btt$ is of great practical and
scientific interest. The reference tool to estimate $\btt$ given data $\bm{y}$ is the \emph{maximum
likelihood estimator} (MLE), which is the vector $\mle$  that minimizes the negative log-likelihood,
given by
\begin{equation} \label{loglik}
   -l(\btt) := \frac{1}{2} \log \big| \bvS(\btt) \big| + \frac{1}{2} (\bm{y} - \bm{\mu})^{T}
  \bvS(\btt)^{-1}
  (\bm{y} - \bm{\mu}) + \frac{n}{2}  \log(2\pi),
\end{equation}
\revs{where $|A|$ denotes the determinant of $A$.} For a detailed discussion of maximum likelihood
and estimation methods, see \cite{Stein1999}. In our discussions here of the log-likelihood, the
mean vector $\bm{\mu}$ will be assumed to be zero, and the constant term will be suppressed. Also,
for notational simplicity, the explicit dependence of $ \bvS $ on $\btt$ will not be indicated in
the rest of this paper. 

From a computational perspective, finding the minimizer of (\ref{loglik}) can be challenging.  The
immediate difficulty is that the linear algebraic operations required to evaluate (\ref{loglik})
grow with cubic time complexity \revs{and quadratic storage complexity}.  Thus, as the data size $n$
increases, direct evaluation of the log-likelihood quickly becomes prohibitively slow or impossible.
As a result of these difficulties, many methods have been proposed that improve both the time and
memory complexity of evaluating or approximating (\ref{loglik}) as well as finding its minimizer in
indirect ways.  Perhaps the oldest systematic methods for the fast approximation of the likelihood
are the ``composite methods,'' which can loosely be thought of in this context as a block-sparse
approximation of $\bvS$ \citep{Vecchia1988, Stein2004, Caragea2007, Katzfuss2017, Katzfuss2017b}. In
a different approach, the methods of matrix tapering \citep{Furrer2006, Kaufman2008, Candas2015} and
Markov random fields \citep{Rue2005, Lindgren2011} use approximations that induce sparsity in $\bvS$
or \revs{$\bvS^{-1}$}, facilitating linear algebra with better scaling that way.  Approximations of
$\bvS$ as a low-rank update to a diagonal matrix have also been applied to this problem
\citep{Cressie2006}, although they can perform poorly in some settings \citep{Stein2014}.  \revs{In
a different approach, by side-stepping the likelihood and instead solving estimating
equations \citep{Godambe1991, Heyde2008}, one can avoid log-determinant computation and perform a
smaller set of linear algebraic operations that are more easily accelerated \citep{Anitescu2011,
Stein2013, Sun2016}.}
The estimating equations approach involves solving a nonlinear system of equations that in the
case of the score equations are given by \revs{setting}
\begin{equation} \label{grad}
  - \nabla l(\btt)_{j} := \frac{1}{2} \text{tr} \left( \bvS^{-1} \bvS_{j} \right) - \frac{1}{2}
  \bm{y}^{T} \bvS^{-1} \bvS_{j} \bvS^{-1} \bm{y}
\end{equation}
to \revs{0 for all $j$}, where here and throughout the paper $\bvS_{j}$ denotes
$\frac{\partial}{\partial \theta_{j}} \bvS(\btt)$. The primary difficulty with computing these
derivatives is that the trace of the matrix-matrix product $ \bvS^{-1} \bvS_{j} $ is prohibitively
expensive to compute exactly. To get around this, \cite{Anitescu2011} proposed using sample average
approximation (SAA), which utilizes the unbiased stochastic trace estimator proposed by
\cite{Hutchinson1990}. For \eqref{grad} it is given by
\begin{equation*} \label{hutch}
  \frac{1}{N_{h}} \sum_{l=1}^{N_{h}} \bm{u}^{T}_{l} \bvS^{-1} \bvS_{j} \bm{u}_{l},
\end{equation*}
where the stochastic $\bm{u}_{l}$ are symmetric Bernoulli vectors (although other options are
available; see \cite{Stein2013} for details). Further, if it is feasible to compute a symmetric
factorization $\bvS = \bm{W} \bm{W}^{T}$, then there may be advantages to using a ``symmetrized''
trace estimator \citep{Stein2013}
\begin{equation*} \label{symhutch}
  \frac{1}{N_{h}} \sum_{l=1}^{N_{h}} \bm{u}^{T}_{l} \bm{W}^{-1} \bvS_{j} \bm{W}^{-T} \bm{u}_{l}.
\end{equation*}
Specifically, writing $\mathcal{I}$ for the expected Fisher information matrix, \cite{Stein2013}
shows that the covariance matrix of these estimates is bounded by $(1+\frac{1}{N_h})\mathcal{I}$,
whereas if we do not symmetrize, it is bounded by
$(1+\frac{(\kappa+1)^2}{4N_h\kappa})\mathcal{I}$, where $\kappa$ is the condition number of the
covariance matrix at the true parameter value.  Since $\kappa\ge 1$, the bound with symmetrization
is at least as small as the bound without it.  Of course, this does not prove that the actual
covariance matrix is smaller, but the results in Section 4 show that the improvement due to
symmetrization can be large.  Moreover, the symmetrized trace estimator reduces the number of linear
solves from two to one if $\bvS^{-1} \by$ is computed as $\bm{W}^{-T} \bm{W}^{-1} \by$, saving
computer time. \revs{Finally, if $\bvS^{-1} \bvS_j$ is itself positive definite, then
probabilistic bounds on the error of the estimator can be controlled independently of the matrix
size $n$ \citep{Roosta2015}}.

Recently, significant effort has been expended to apply the framework of hierarchical matrices
\citep{Hackbusch1999, Grasedyck2003, Hackbusch2015}---which utilize the low-rank block structure of
certain classes of special matrices to obtain significantly better time and storage complexity for
common operations---to the problem of Gaussian process computing \citep{Borm2007, Ambikasaran2016,
Minden2016, Litvinenko2017, Chen2017}.
%
We follow in this vein here, extending some of these ideas by using the specific class of
hierarchical matrices referred to as hierarchical off-diagonal low-rank (HODLR) matrices
\citep{Ambikasaran2013}. Unlike some of the methods described earlier, approaches to approximating
$l_{H}$ with hierarchical matrices have the benefit of being able to directly compute the
log-determinant of $\bvS$ and the linear solve $\bvS^{-1} \bm{y}$\revs{, which would not be possible
if one were to use matrix-free methods like the Fast Multipole Method (FMM) \citep{Greengard87} or
circulant embedding \citep{Anitescu2011}.} Letting $\hbvS$ denote a hierarchical approximation of $\bvS$ here and throughout
the paper, we give the approximated log-likelihood as
\begin{equation} \label{hloglik} - l_{H}(\btt) := \frac{1}{2} \log \big| \hbvS \big| +
  \frac{1}{2} \bm{y}^{T} \hbvS^{-1} \bm{y}. 
\end{equation} 
While the issue of how well $\hbvS$ approximates $\bvS$ remains, the applications of H-matrices to
maximum likelihood cited above
have demonstrated that such approximations for covariance matrices can yield good estimates of
$\mle$.

Most investigations into the use of hierarchical matrices in this area focus exclusively on
the computation of $l_{H}(\btt)$, and not its first- and second-order derivatives.  If
the goal is to carry out minimization of $-l_{H}(\btt)$, however, access to such information would
significantly reduce the number of iterations required to approximate $\mle$ \citep{Nocedal99}. Part
of the difficulty is that the matrix product $\hbvS^{-1} \hbvS_{j}$ is still expensive to
compute for hierarchical matrices, and so the trace term in the score equations remains challenging
to obtain quickly. As a result, stochastic methods for estimating the trace associated with
the gradient of $l_{H}$ are necessary in order to maintain good complexity. An alternative method to
the Hutchinson estimator is suggested by \cite{Minden2016}, who utilize the peeling algorithm of
\cite{Lin2011} to assemble a hierarchical approximation of $\hbvS^{-1} \hbvS_{j}$ and obtain a more
precise estimate for its trace. In this setting especially, however, this is done at the cost of
substantially higher overhead than occurs with the Hutchinson estimator. As is demonstrated in this paper, the
symmetrized Hutchinson estimator is reliable enough for the purpose of numerical optimization. 

One important choice for the construction of hierarchical approximations is the method for
compressing low-rank off-diagonal blocks; we refer readers to \cite{Ambikasaran2016} for a
discussion. In this work, we advocate the use of the \Nystrom approximation \citep{Williams2001,
Drineas2005, Chen2017}. The \Nystrom approximation of the block of $\bvS$ corresponding to indices $I$
and $J$ is given by
\begin{equation} 
  \label{nys} \hbvS_{I, J} := \bvS_{I, P} \bvS_{P, P}^{-1} \bvS_{P, J},
\end{equation} 
where the indices $P$ are for $p$ many \emph{landmark points} that are chosen from the dataset.  As
can be seen, this formulation is less natural to use in an adaptive way than, for example, a
truncated singular value decomposition or other common fast approximation methods. Unlike most
adaptive methods  \citep{Griewank2008evaluating}, however,  it is differentiable with respect to the
parameters $\btt$, a property that is essential to the approach advocated here. As a result of this
unique property, the derivatives of $\hbvS(\btt)$ are both well defined and computable in
quasilinear complexity if its off-diagonal blocks are constructed with the \Nystrom approximation,
\revs{so that the second term in the hierarchically approximated analog of (\ref{grad}) can be computed
exactly.}

In this paper, we discuss an approach to minimizing (\ref{hloglik}) that combines the HODLR matrix
structure from \cite{Ambikasaran2013}, the \Nystrom approximation of \cite{Williams2001}, and the
sample average (SAA) trace estimation from \cite{Anitescu2011} and \cite{Stein2013} in its
symmetrized form.  As a result, we obtain stable and optimization-suitable stochastic estimates
for the gradient and Hessian of (\ref{hloglik}) in quasilinear time and storage complexity at a
comparatively low overhead.  Combined with the exact derivatives of $\hbvS$, the symmetrized
stochastic trace estimators are demonstrated to yield gradient and Hessian approximations with
relative numerical error below $0.03\%$ away from the MLE for a manageably small number of
$\bm{u}_{l}$ vectors.  As well as providing tools for optimization, the exact derivatives and
stabilized trace estimation mean that the observed and expected Fisher information matrix may
be computed in quasilinear complexity, serving as a valuable tool for estimating the covariance
matrix of $\mle$.

\subsection{Comparison with existing methods}

Like the works of \cite{Borm2007}, \cite{Anitescu2011}, \cite{Stein2013}, \cite{Minden2016},
\cite{Litvinenko2017}, and \cite{Chen2017}, we attempt to provide an approximation of the
log-likelihood that can be computed with good efficiency but whose minimizers closely resemble those
of the exact likelihood. The primary distinction between our approach and other methods is that we
construct our approximation with an emphasis on having mathematically well-defined and
computationally feasible derivatives.  By computing the exact derivative of the approximation of
$\bvS$, given by $\hbvS_j=\frac{\partial}{\partial \theta_{j}} \hbvS$, instead of an independent
approximation of the derivative of the exact $\bvS$, which might be denoted by
$\widetilde{\frac{\partial}{\partial \theta_{j}} \bvS}$ to emphasize that one is approximating the
derivative of the exact matrix $\bvS(\btt)$,
we obtain a more coherent framework for thinking about both optimization and error
propagation in the derivatives of $l_{H}$.  As an example of the practical significance of this
distinction, for a scale parameter $\theta_{0}$ and covariance matrix parameterized with $\hbvS =
\theta_{0} \hbvS'$, the derivative $\frac{\partial}{\partial \theta_{0}} \hbvS$ will be numerically
identical to $\hbvS'$, making $\hbvS^{-1} \hbvS_{j}$ an exact rescaled identity matrix.  As a result,
the stochastic Hutchinson trace estimator of that matrix-matrix product for scale
parameters is 
exact to numerical precision with a single $\bm{u}_{l}$, which will be reflected
in the numerical results section. To our knowledge, such a guarantee cannot be made if $\hbvS_{j}$ is
constructed as an approximation of the exact derivative $\bvS_{j}$. 
Moreover, none of the
methods mentioned above discuss computing Hessian information of any kind.

Our approach achieves quasilinear complexity both in evaluating the log-likelihood and in computing
accurate and stable stochastic estimators for the gradient and Hessian of the approximated
log-likelihood. This comes at the cost of abandoning a priori controllable bounds on pointwise
precision of the hierarchical approximation of the exact covariance matrix, a less accurate trace
estimator than has been achieved with the peeling method \citep{Minden2016, Lin2011}, and
sub-optimal time and storage complexity \citep{Chen2017}.  Nevertheless, we consider this to be a
worthwhile tradeoff in some applications, and we demonstrate in the numerical results section that
despite the loss of control of pointwise error in the covariance, we can compute estimates for MLEs
and their corresponding uncertainties from the expected Fisher matrix that agree well with exact
methods.  Further, by having access to a gradient and Hessian, we have many options for
numerical optimization and are able to perform it reliably and efficiently.

\section{HODLR matrices, derivatives, and trace estimation}

In this study, we approximate $\bvS$ with the HODLR format \citep{Ambikasaran2013}, which has an
especially simple and tractable structure given by
\begin{equation} \label{HODLR}
  \begin{bmatrix}
    \bm{A}_{1} && \bm{U}\bm{V}^{T} \\
    \bm{V} \bm{U}^{T} && \bm{A}_{2}
  \end{bmatrix},
\end{equation}
where the matrices $\bm{U}$ and $\bm{V}$ are of dimension $n \times p$ and $\bm{A}_{1}$ and
$\bm{A}_{2}$ are either dense matrices or are of the form of (\ref{HODLR}) themselves.  A matrix of
size $n \times n $ can be split recursively into block $2 \times 2$ matrices in this way $\lfloor
\log_{2}(n) \rfloor$ times, although in practice it is often divided fewer times than that. The
diagonal blocks of a HODLR matrix are often referred to as the \emph{leaves}, referring to the fact
that a tree is implicitly being constructed.  

Symmetric positive definite HODLR matrices admit an exact symmetric factorization $\hbvS = \bm{W}
\bm{W}^{T}$ that can be computed in $O(n \log^2 n)$ complexity if $p$ is fixed and the level grows
with $O(\log n)$ \citep{Ambikasaran2014a}.
For a matrix of level $\tau$, $\bm{W}$ takes the form 
\begin{equation*}
  \bm{W} = \overline{\bm{W}} \prod_{k=1}^{\tau} \left\{ \mathbb{I} + \overline{\bm{U}}_{k}
  \overline{\bm{V}}_{k}^{T} \right\},
\end{equation*}
where $\overline{\bm{W}}$ is a block-diagonal matrix of the symmetric factors of the leaves
$\bm{L}_{k}$ and each $\mathbb{I} + \overline{\bm{U}} \; \overline{\bm{V}}^{T}$ is a block-diagonal
low-rank update to the identity.

If the rank of the off-diagonal blocks is fixed at $p$ and the level grows with $O(\log n)$, then
the log-determinant and linear system computations can both be performed at $O(n \log n)$ complexity
by using the symmetric factor \citep{Ambikasaran2014a}. With these tools, we may evaluate the
approximated Gaussian log-likelihood given in (\ref{hloglik}) exactly and directly. We note that the
assembly of the matrix and its factorization are parallelizable and that many of the computations in
the numerical results section are done in single-node multicore parallel. \revs{With that said,
however, we also point out that effective parallel implementations of hierarchical matrix
operations are challenging, and that the software suite that is companion to this paper is not
sufficiently advanced that it benefits from substantially more threads or nodes than a
reasonably powerful workstation would provide.}

\subsection{The \Nystrom approximation and gradient estimation}
As mentioned in the Introduction, the method we advocate here for the low-rank compression of
off-diagonal blocks is the \Nystrom approximation \citep{Williams2001}, a method recently applied to
Gaussian process computing and hierarchical matrices \citep{Chen2017}.  Unlike the multiple common
algebraic approximation methods that often construct approximations of the form $\bm{U} \bm{V}^{T}$
by imitating early-terminating pivoted factorization \citep{Ambikasaran2016}, for example the
commonly used adaptive cross-approximation (ACA) \citep{Bebendorf2000, Rjasanow02},  the \Nystrom
approximation constructs approximations that are continuous with respect to the parameters $\btt$ in
a nonadaptive way.  
Another advantage of this method is that an approximation $\hbvS$ assembled with the \Nystrom
approximation is guaranteed to be positive definite if $\bvS$ is \citep{Chen2017},
avoiding the common difficulty of guaranteeing that a hierarchical approximation $\hbvS$ of positive
definite $\bvS$ is itself positive definite \citep{Bebendorf2007, Xia2010, Chen2017}.

The cost of choosing the \Nystrom approximation for off-diagonal blocks instead of a method like the
ACA is that we cannot easily adapt it locally to a prescribed accuracy. That is, 
constructing a factorization with a precision $\eps$ so that
$
  || \bvS_{I, J} - \hbvS_{I, J} || < \eps \||\bvS_{I, J}||
$
is \revs{not generally possible} since we must choose the number and locations of the landmark
points $P$ before starting computations.  \revs{Put in other terms, the rank of every off-diagonal
block approximation  is the same, and there are no guarantees for the pointwise quality of
that fixed-rank approximation.}

The primary appeal of this approximation for our purposes, however, is that its derivatives exist
and are readily computable by the product rule, given by
\begin{equation} \label{deriv}
  \hbvS_{j,(I,J)} = 
  \bvS_{j, (I, P)} \bvS_{P,P}^{-1} \bvS_{P,J} -
  \bvS_{I, P} \bvS_{P,P}^{-1} \bvS_{j, (P,P)} \bvS_{P,P}^{-1} \bvS_{P,J} +
  \bvS_{I, P} \bvS_{P,P}^{-1} \bvS_{j, (P,J)},
\end{equation}
where, for example, $\bvS_{j,(I,J)}=\frac{\partial}{\partial \btt_{j}} \bvS_{I,J}$, with parentheses
and capitalization used to emphasize subscripts denoting block indices. Fortunately, all three terms
in (\ref{deriv}) are the product of $n \times p$ and $p \times p$ matrices. Since the diagonal
blocks of $\hbvS_{j}$ are trivially computable as well, the exact derivative of
$\hbvS$ can be represented as a HODLR matrix with the same shape as $\hbvS$ except that now the
off-diagonal blocks are sums of three terms that look like truncated factorizations of the form
$\bm{U} \bm{S} \bm{V}^{T}$, where $\bm{S} \in \mathbb{R}^{p \times p}$. In practice, assembling and
storing $\hbvS_{j}$ take only about twice as much time and memory as required for $\hbvS$. Thus, one
can compute terms of the form
$
  \bm{y}^{T} \hbvS^{-1} \hbvS_{j} \hbvS^{-1} \bm{y}
$
exactly and in quasilinear time and storage complexity, \revs{providing an exact method for obtaining the
second term in the gradient of $l_H$}.

We now combine our results from the preceding section with the symmetrized trace estimation discussed
in the Introduction and obtain a stochastic gradient approximation for $-l_{H}$ that can be computed
in quasilinear complexity: 
\begin{equation*} \label{stoch_grad}
  \widehat{\nabla -l_{H}(\btt)}_{j} = \frac{1}{2 N_{h}} \sum_{l=1}^{N_{h}} \bm{u}^{T}_{l}
  \bm{W}^{-1} \hbvS_{j} \bm{W}^{-T} \bm{u}_{l} - \frac{1}{2}\bm{y}^{T} \hbvS^{-1} \hbvS_{j}
  \hbvS^{-1} \bm{y}.
\end{equation*}
Both the symmetrized trace estimator, facilitated by the symmetric factorization, and the exact
derivatives, facilitated by the \Nystrom approximation, are important to the performance of this
estimator. While the benefit of the latter is clear, the decreased variance and faster computation
time are nontrivial benefits as well. The numerical results section has a brief demonstration of
these benefits.

\subsection{Stochastic estimation of information matrices}
Being able to efficiently and effectively estimate trace terms involving the derivatives $\hbvS_{j}$
also facilitates accurate estimation of the expected Fisher information matrix, which has terms
given by 
\begin{equation*} \label{exac_efisher}
  \mathcal{I}_{j,k} = \frac{1}{2} \text{tr} \left(\hbvS^{-1} \hbvS_{j} \hbvS^{-1} \hbvS_{k} \right).
\end{equation*}
Using the same symmetrization approach as above, we may compute stochastic estimates of these terms
with the unbiased and fully symmetrized estimator
\begin{equation} \label{stoch_efisher}
  \revs{
  \widehat{\mathcal{I}}_{j,k} = 
  \frac{1}{4 N_{h}} \sum_{l=1}^{N_{h}} \bm{u}^{T}_{l} \bm{W}^{-1} \left( \hbvS_{j} + \hbvS_{k}
  \right) \hbvS^{-1} \left( \hbvS_{j} + \hbvS_{k} \right) \bm{W}^{-T} \bm{u}_{l} -
  \frac{1}{2} \widehat{\mathcal{I}}_{j,j} - \frac{1}{2} \widehat{\mathcal{I}}_{k,k}, \; j \neq k
  }
\end{equation}
Since the diagonal elements of $\widehat{\mathcal{I}}$ can be computed first in a simple and
trivially symmetric way, this provides a fully symmetrized method for estimating $\mathcal{I}_{j,k}$.
Further, computing the estimates in the form of (\ref{stoch_efisher}) does not require any more
matvec applications of derivative matrices than would the more direct estimator that computes terms
as
$
  \bm{u}^{T}_{l} \bm{W}^{-1} \hbvS_{j}  \hbvS^{-1}  \hbvS_{k} \bm{W}^{-T} \bm{u}_{l},
$
since the terms in (\ref{stoch_efisher}) look like $\bm{u}^{T} \bm{A} \bm{A}^{T} \bm{u} = ||\bm{A}^{T}
\bm{u}||^{2}$, if we recall that the innermost solve $\hbvS^{-1} \by$ is computed by sequential
solves $\bm{W}^{-T} \bm{W}^{-1} \by$. As a result, each term in (\ref{stoch_efisher}) still 
requires only one full solve with $\hbvS$ and two derivative matrix applications. \revs{In
circumstances where one expects $\hbvS^{-1} \hbvS_j \hbvS^{-1} \hbvS_k$ to itself be positive
definite, then the stronger theoretical results of \cite{Roosta2015} would apply, giving a high
level control over the error of the stochastic trace estimator.}

Having efficient and accurate estimates for $\mathcal{I} \left( \btt \right)$ is helpful
because they can be used for confidence intervals, since asymptotic theory suggests \citep{Stein1999}
that if the smallest eigenvalue of $\mathcal{I}$ tends to infinity as the sample size increases,
then we can expect that
\begin{equation*} \label{mledist}
  \mathcal{I}(\hat{\btt})^{1/2} (\hat{\btt}-\btt_{\mbox{true}}) \to_D N(0,\mathbb{I}).
\end{equation*}

The Hessian of $-l_{H}$, which requires computing second derivatives of $\hbvS$, is useful for both
optimization and inference.  The exact terms of the Hessian $(H l_{H})_{j,k}$ are given by 
\begin{equation}\label{eq:loglikHess}
  \frac{1}{2} \left( 
  -\text{tr} \left( \hbvS^{-1} \hbvS_{k} \hbvS^{-1} \hbvS_{j} \right) 
  +
  \text{tr} \left(  \hbvS^{-1} \hbvS_{jk} \right)
  \right)
  - 
  \frac{1}{2} \by^{T} \left(\frac{\partial}{\partial \theta_{k}} \hbvS^{-1} \hbvS_{j} \hbvS^{-1} \right) \by,
\end{equation}
where
\begin{equation*}
  \frac{\partial}{\partial \theta_{k}} \hbvS^{-1} \hbvS_{j} \hbvS^{-1} = 
  -
  \hbvS^{-1} \hbvS_{k} \hbvS^{-1} 
  \hbvS_{j} \hbvS^{-1} 
  +
  \hbvS^{-1} \hbvS_{jk} \hbvS^{-1} 
  -
  \hbvS^{-1} \hbvS_{j} \hbvS^{-1} 
  \hbvS_{k} \hbvS^{-1} .
\end{equation*}
Continuing further with the symmetrization approach, we obtain the unbiased fully symmetrized
stochastic estimator of the first two terms of (\ref{eq:loglikHess}) given by
\begin{equation*} \label{ddk_trace}
   \frac{1}{2N_h} \sum_{l=1}^{N_h} \bm{u}_{l}^{T} \bm{W}^{-1} \hbvS_{jk} \bm{W}^{-T} \bm{u}_{l} -
   \widehat{\mathcal{I}}_{j,k},
\end{equation*}
where $\widehat{\mathcal{I}}_{j,k}$ is the $j,k$th term of the estimated Fisher information matrix.

Since the third and fourth terms in (\ref{eq:loglikHess}) can be computed exactly, replacing the two
trace terms in \eqref{eq:loglikHess} with the stochastic trace estimator provides a stochastic
approximation for the Hessian of $-l_{H}$. The computation of the second partial derivatives
$\hbvS_{jk}$ is a straightforward continuation of Equation (\ref{deriv}) and will again result in
the sum of a small number of HODLR matrices with the same structure. 
The overall effort of estimating the Hessian of (\ref{hloglik}) with our approach will thus also
have $O(n \log^2 n)$ complexity, providing extra tools for both optimization and covariance matrix
estimation, since in some circumstances the observed information is a preferable estimator to the
expected information \citep{Efron78}.   We note, however, that our approximation of the expected
Fisher information is guaranteed to be positive semidefinite if one uses the same $\bm{u}_{l}$
vectors for all components of the matrix, whereas our approximation of the Hessian may not be
positive semidefinite at the MLE. Exact expressions for $\hbvS_{jk}$ are
given in the Appendix.

\section{Numerical results}

We now present several numerical experiments to demonstrate both the accuracy and scalability of the
proposed approximation to the log-likelihood for Gaussian process data. 
In the sections below, we demonstrate the effectiveness of this method using two parameterizations of
the Mat\'ern covariance, which is given in its most standard form by
\begin{equation} \label{matern}
  K(\bx,\by;\btt,\nu) := \theta_{0} \mathcal{M}_{\nu}\left( \frac{||\bx-\by||}{\theta_{1}} \right).
\end{equation}
Here, $\mathcal{M}_{\nu}$ is the Mat\'ern correlation function, given by
\begin{equation*}
  \mathcal{M}_{\nu}(x) := \big(2^{\nu-1} \Gamma(\nu) \big)^{-1} \big( \sqrt{2 \nu} x \big)^{\nu}
  \mathcal{K}_{\nu} \big( \sqrt{2 \nu} x \big) ,
\end{equation*}
and $\mathcal{K}_{\nu}$ is the modified Bessel function of the second kind.  The quantity
$\theta_{0}$ is a scale parameter, $\theta_{1}$ is a range parameter, and $\nu$ is a smoothness
parameter that controls the degree of differentiability if the process is differentiable or,
equivalently, the high-frequency behavior of the spectral density \citep{Stein1999}.

As well as demonstrating the similar behavior of the exact likelihood to the approximation we
present, this section explores the important algorithmic choices that can be tuned or
selected. These are (i) the number of block-dyadic divisions of the matrix (the \emph{level}
of the HODLR matrix) and (ii) the globally fixed \emph{rank} of the off-diagonal blocks. These
choices are of particular interest in addressing possible concerns that the resulting
estimate from minimizing $-l_{H}$ may be sensitive to the choices of these parameters and that choosing
reasonable a priori values may be difficult, although we demonstrate below that the method is
relatively robust to these choices.

In all the numerical simulations and studies described below, unless otherwise stated, the fixed
rank of off-diagonal blocks has been set at $72$, the level at $\lfloor \log_{2} n \rfloor - 8$,
which results in diagonal blocks (leaves) with sizes between $256$ and $512$\revs{, and 35 random
vectors (fixed for the duration of an optimization routine) are used for stochastic trace
estimation}. The ordering of the observations/spatial locations is done through the traversal of a
K-D tree as in \cite{Ambikasaran2016}, which is a straightforward extension to the familiar
one-dimensional tree whose formalism is dimension-agnostic \revs{and resembles a multidimensional
analog to sorting. By ordering our points in this way, we increase the degree to which off-diagonal
blocks correspond to the covariance between groups of well-separated points, which is a critical
requirement and motivation for the hierarchical approximation of $\bvS$.}

Writing $\btt_{-0}$ for all components of $\btt$ other than the scale parameter $\theta_{0}$, the
minimizer of $-l_H(\theta_0,\btt_{-0})$ for fixed $\btt_{-0}$ as a function of $\theta_0$ is
$\hat{\theta}_0(\btt_{-0}) = n^{-1} \by^{T} \bvS(1, \btt_{-0})^{-1} \by$.  Thus, the negative
log-likelihood can be minimized by instead minimizing the negative \emph{profile log-likelihood},
given by
\begin{equation*} \label{hloglik_profile}
 - l_{H, \text{pr}}(\btt_{-0}) = -l_H(\hat{\theta}_0(\btt_{-0}),\btt_{-0}) = \frac{1}{2} \log | \hbvS(1,\btt_{-0}) | + \frac{n}{2} \log
 \left( \by^{T}
  \hbvS(1,\btt_{-0})^{-1} \by \right),
\end{equation*}
which reduces the dimension of the minimization problem by one parameter. Its derivatives and trace
estimators follow similarly as for the full log-likelihood. 

All the computations shown in this section were performed on a standard workstation with an Intel
Core i7-6700 processor \revs{with $8$ threads and} $40$ GB of RAM, and the assembly of $\hbvS$,
$\hbvS_{j}$, and $\hbvS_{jk}$ was done in multicore parallel \revs{with $6$ threads}, as was the
factorization of $\hbvS = \bm{W} \bm{W}^{T}$.  A software package written in the Julia programming
language \citep{Bezanson17},
\texttt{KernelMatrices.jl}, which provides the source code to perform all the computations described
in this paper as well as reproduce the results in this section, is available at
\texttt{bitbucket.org/cgeoga/kernelmatrices.jl.}

\subsection{Quasilinear scaling of the log-likelihood, gradient, and Hessian}

\begin{figure}[!ht] 
  \begin{center}
    \begin{tabular}{ccc}
      \includegraphics[width=.32\linewidth]{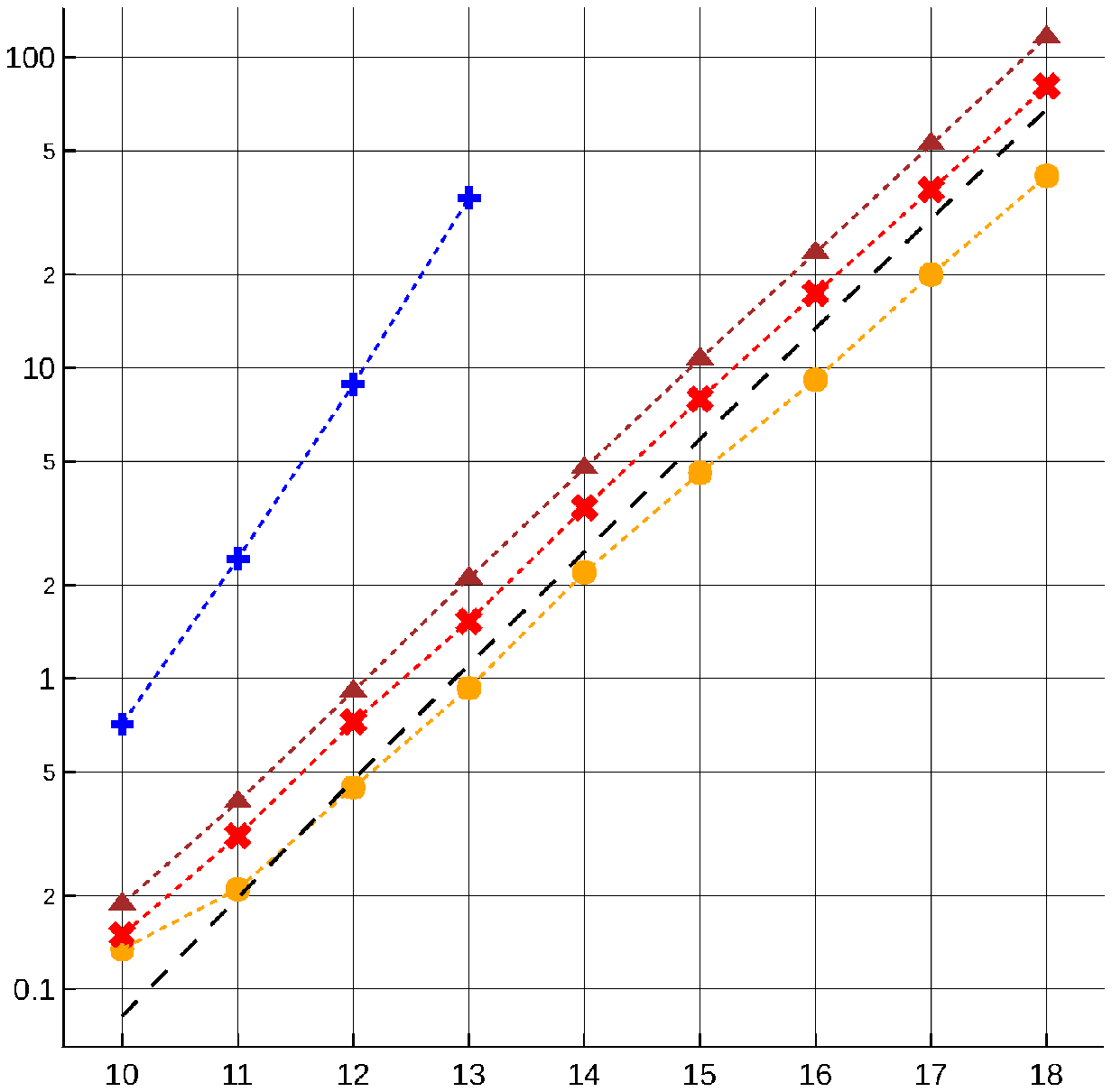} &
      \includegraphics[width=.32\linewidth]{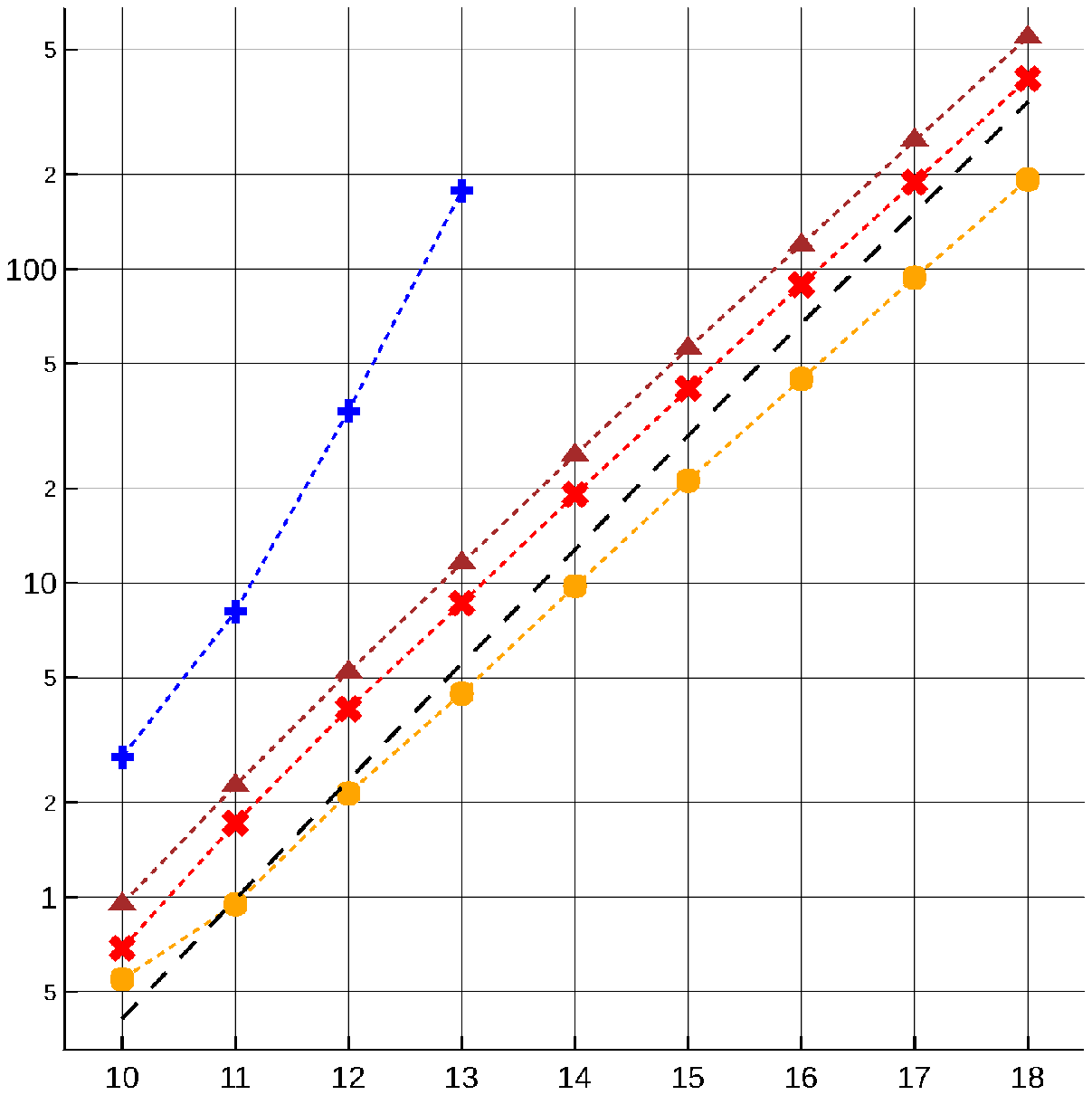} &
      \includegraphics[width=.32\linewidth]{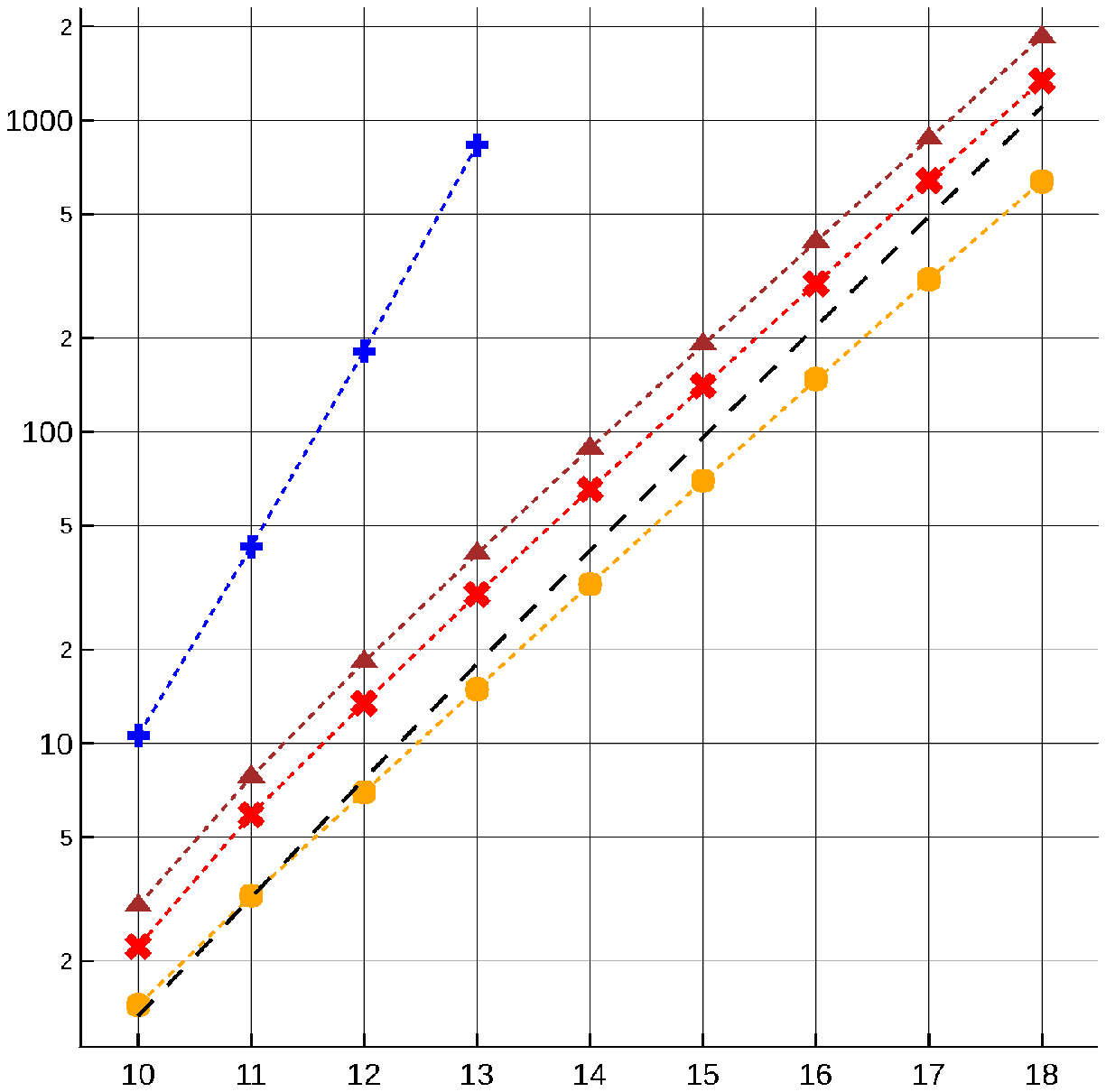} \\
      (a) & (b) & (c) \\
    \end{tabular}
    \caption{Times taken \revs{(in seconds)} to call the likelihood (a), gradient (b), and Hessian (c)
      exactly (plus) and their HODLR equivalents for fixed off-diagonal rank 32 (circle), 72 (x),
      and 100 (triangle) for sizes $2^k$ with $k$ from $10$ to $18$ (horizontal axis). Theoretical lines
      corresponding to $O(n \log^{2} n)$ are added to each plot to demonstrate the scaling.
    \label{figure:timecall} }
  \end{center}
\end{figure}

To demonstrate the scaling of the approximated log-likelihood, its stochastic gradient, and
its Hessian, we show in Figure \ref{figure:timecall} the average time taken to evaluate those functions
for the two-parameter (fixing $\nu$) full Mat\'ern log-likelihood (as opposed to profile likelihood) for
sizes $n=2^{k}$ for $k$ ranging from $10$ to $18$, including also the time taken to call the exact
functions for $k \leq 13$.

As can be seen in Figure \ref{figure:timecall}, the scaling of all three operations for the
approximated log-likelihood follow the expected quasilinear $O(n \log^{2} n)$ growth, if not scaling even
slightly better. For practical applications, the total time required to compute these values
together will be slightly lower than the sum of the three times plotted here because of repeated
computation of the derivative matrices $\hbvS_{j}$, repeated linear solves of the form $\hbvS^{-1}
\by$, and other micro-optimizations that combine to save computational effort.

\subsection{Numerical accuracy of symmetrized trace estimation}

\begin{figure}[!ht]
  \begin{center}
    \begin{tabular}{cc}
      \includegraphics[width=.49\linewidth]{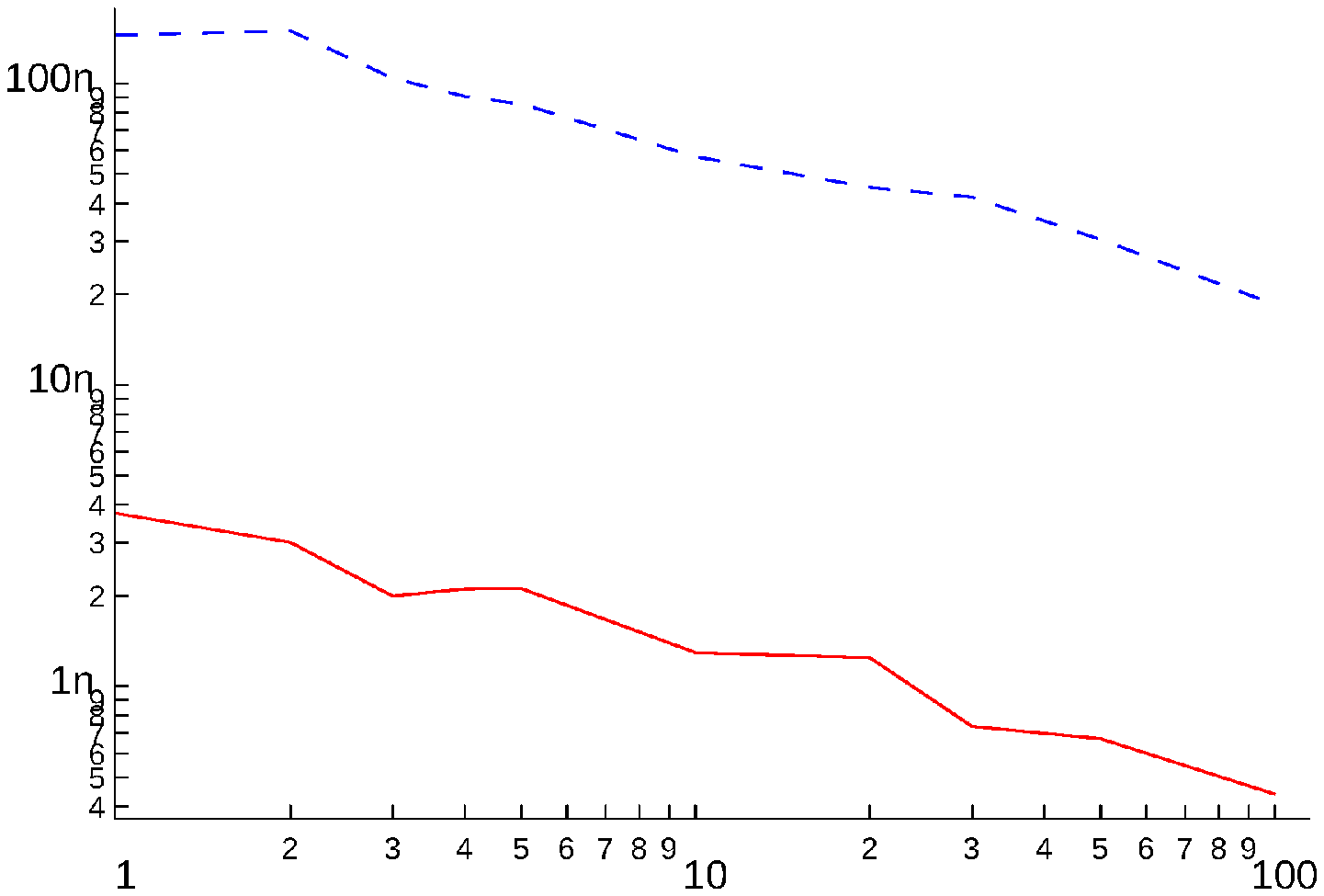} &
      \includegraphics[width=.49\linewidth]{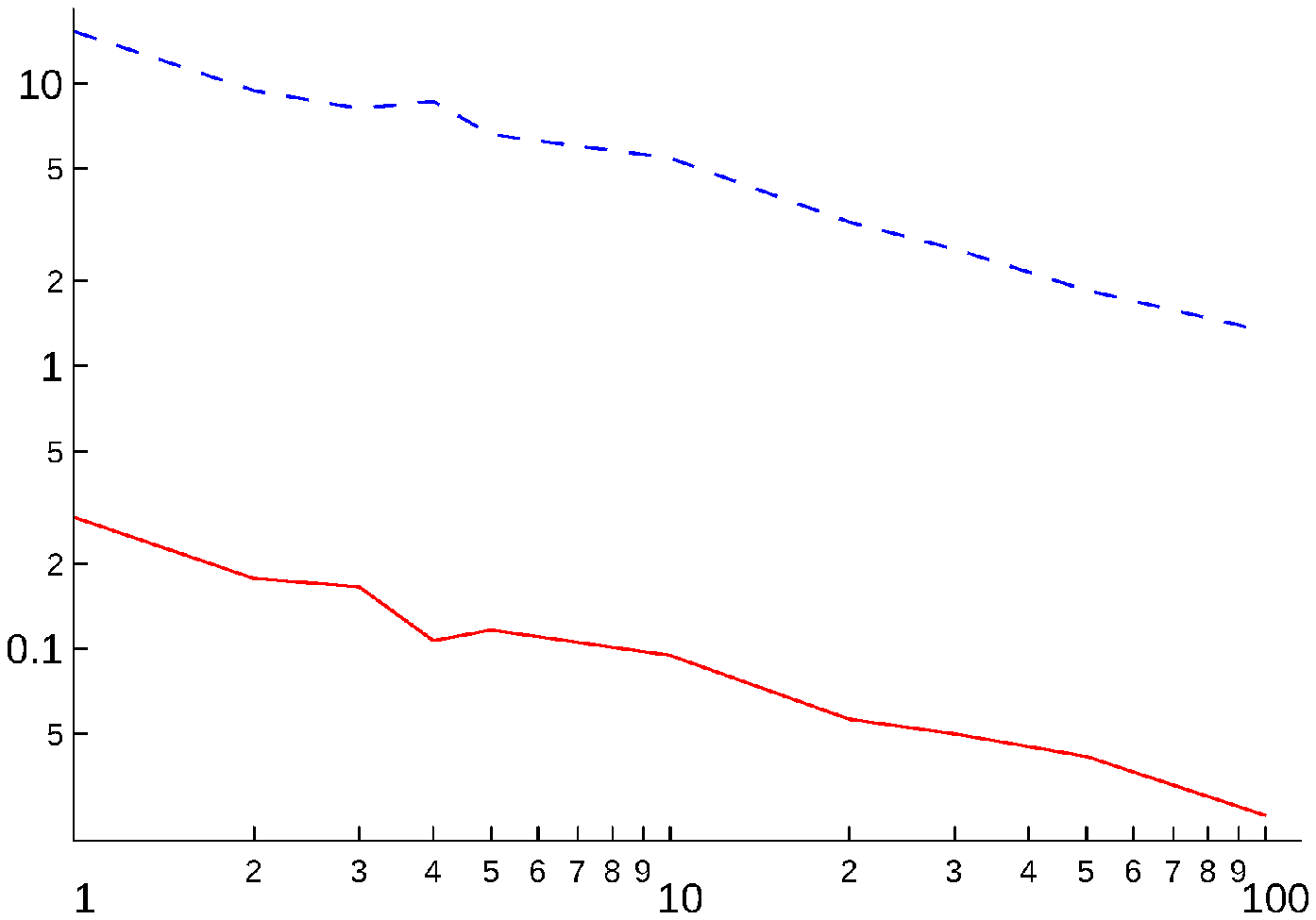}  \\
      (a) & (b) \\
    \end{tabular}
    \caption{Standard deviation of 50 stochastic trace estimates for $\hbvS^{-1} \hbvS_{j}$ for the
    scale parameter (a) and range parameter (b) of the Mat\'ern covariance. As can be seen,
    nonsymmetrized estimates (dashed) have standard deviations more than one order of
    magnitude larger than their symmetrized counterparts (solid). In (a), we use the notation
    $1n = 10^{-9}$. \label{figure:trace_plots}}
  \end{center}
\end{figure}

To demonstrate the benefit of the symmetrized stochastic trace estimation described in the
Introduction, we simulate a process at random locations in the domain $[0, 100]^2$ for Mat\'ern
covariance with parameters $\theta_{0} = 3$, $\theta_{1} = 40$, and $\nu$ fixed at $1$ and then
compare the standard and symmetrized trace estimators for  $\hbvS^{-1}\hbvS_{j}$ associated with
$\theta_{0}$ and $\theta_{1}$. We choose a large range parameter with respect to the edge length of
the domain in order to demonstrate that even for poorly conditioned $\hbvS$, the symmetrized trace
estimator performs well. As the standard deviations shown in Figure \ref{figure:trace_plots}
demonstrate, the symmetrized trace estimator reduces the standard deviation of estimates by more
than a factor of 10. As remarked in the Introduction,  for the scale parameter $\theta_0$,
$\hbvS^{-1}\hbvS_{j}$ is a multiple of the identity matrix, so there is no stochastic error even for
one $\bm{u}_{l}$, and the errors reported here are purely numerical. 

\subsection{Effect of method parameters on likelihood surface}

\begin{figure}[!ht] 
  \begin{center}
    \includegraphics[width=\textwidth]{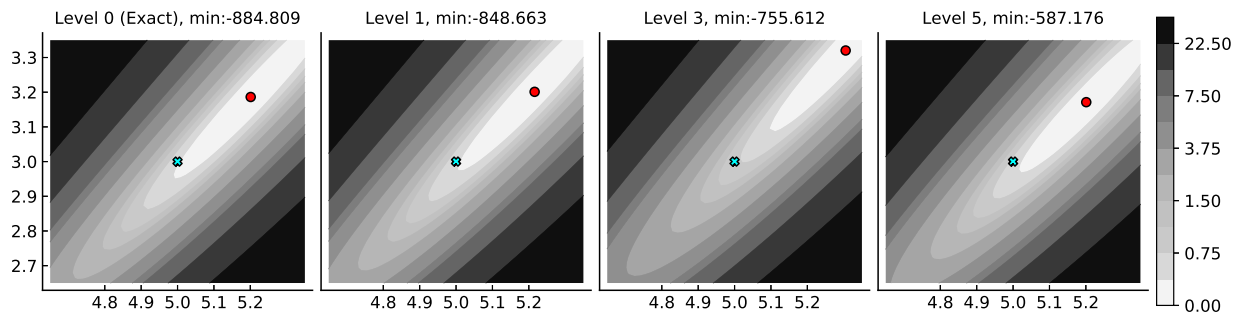}
    \caption{Centered log-likelihood surface for $n=2^{12}$ data points randomly sampled on the box
    $[0,100]^2$ with Mat\'ern covariance with parameters $\theta_{0}=3$, $\theta_{1}=5$, and
    $\nu=1$, with $\nu$ assumed known.  The blue x is the ``true'' parameters of the simulation,
    and the red circle is the minimizer. The value at the minimizer is shown with the level in the title.
    \label{figure:lvl_surf} }
  \end{center}
\end{figure}

Using the Mat\'ern covariance function, we simulate $n=2^{12}$ observations of a random field with
Mat\'ern covariance function with parameters $\theta_{0} = 3.0$, $\theta_{1} = 5.0$, and fixed
$\nu=1$ at random locations on the box $[0, 100]^2$.  In Figure \ref{figure:lvl_surf}, which shows
the log-likelihood surfaces for various levels with the minimizer of each subtracted off, we see the
minimal effect on the location of the minimizer caused by varying the level of the HODLR
approximation with off-diagonal rank fixed at $64$ for nonexact likelihoods \revs{so that the fixed
rank of off-diagonal blocks does not exceed their size at level $5$. Note that each graphic is on
the same color scale and has the same contour levels, so that the only noticeable change between
levels is an additive translation.  For a similar comparison for different fixed off-diagonal
ranks that yields the same interpretation, see the supplementary material.} 

\subsection{Numerical accuracy of stochastic derivative approximations}

To explore the asymptotic behavior of the minimizers of $-l_{H}$, we now switch to a different
parameterization of the Mat\'ern covariance inspired by \cite{Stein1999} and \cite{Zhang2004}, given
by
\begin{equation} \label{smatern}
  K_{s}(\bx,\by;\btt,\nu) := \theta_0 \left( \frac{\theta_1}{2\sqrt{\nu}} ||\bx - \by|| \right)^\nu
  \mathcal{K}_\nu \left( \frac{2\sqrt{\nu}}{\theta_1} ||\bx - \by|| \right).
\end{equation}
The advantage of this parameterization is that, unlike in (\ref{matern}), it clearly separates
parameters that can be estimated consistently as the sample size grows on a fixed domain from those
that cannot \citep{Stein1999, Zhang2004, Zhang2005}. Specifically, for bounded domains in 3 or fewer
dimensions, results on equivalence and orthogonality of Gaussian measures suggest that both
$\theta_0$ and $\nu$ can be estimated consistently under the parameterization (\ref{smatern}),
whereas $\theta_0$ definitely cannot be estimated consistently under (\ref{matern}).  The range parameter
$\theta_{1}$ cannot be estimated consistently under either parameterization. 

To demonstrate the accuracy of the stochastic gradient and Hessian estimates of $-l_{H}$ with this
covariance function, we compare them with the exact gradient and Hessian of the approximated
log-likelihood computed directly for a variety of small sizes.  For the specific setting of the
computations, we simulate Gaussian process data at random locations on the domain $[0, 100]^2$ with
parameters $\btt = (3, 5)$ and $\nu$ fixed at $1$ to avoid the potentially confounding numerical
difficulties of computing $\frac{\partial}{\partial \nu} \mathcal{K}_{\nu}(x)$. 

To explore the numerical accuracy of the stochastic approximations, we consider both estimates at
the computed MLE $\mle$ and at a potential starting point for optimization, which was chosen to be
$\btt_{\text{init}} = (2,2)$. For both cases, we consider the standard relative precision metric.
We note, however, that if $-l_H$ were exactly minimized, its gradient would be exactly 0 and the
relative precision would be undefined. Thus, we also consider the alternative measures of accuracy
at the evaluated MLE given by 
\begin{equation} \label{grad_prec_metric}
  \eta_{g} := || \widehat{\nabla l_{H} (\btt)} - \nabla l_{H} (\btt) ||_{\mathcal{I} (\btt)^{-1}}
\end{equation}
for the gradient and 
\begin{equation} \label{hess_prec_metric}
  \eta_{\mathcal{I}} := \text{tr} \left\{
  \big( \widehat{\mathcal{I}} (\btt)      - \mathcal{I} (\btt) \big)
  \big( \mathcal{I} (\btt)^{-1} - \widehat{\mathcal{I}} (\btt)^{-1} \big)
\right\} ^{1/2}
\end{equation}
for the expected Fisher matrix. These measures of precision are stable at the MLE and are invariant
to all linear transformations; $\eta_{\mathcal{I}}$ is a natural metric for positive definite
matrices in that, for fixed positive definite $\mathcal{I}$, it tends to infinity as a sequence of
positive definite $\widehat{\mathcal{I}}$ tends to a limit that is only positive semidefinite.
Using $\eps$ to denote the standard relative precision, Tables \ref{tab:stch_prec_far} and
\ref{tab:stch_prec_mle} summarize these results.

\begin{table}[!ht]
  \begin{center}
    \caption{Average relative precision (on $\log_{10}$ scale) of the stochastic gradient and Hessian of
    the approximated log-likelihood for 5 simulations with $\btt = (2,2)$.
    \label{tab:stch_prec_far} }
    \begin{tabular}{|c|c|c|c|c|}
      \hline
                                            & $n=2^{10}$ & $n=2^{11}$ & $n=2^{12}$ & $n=2^{13}$  \\
      \hline
      $\eps_{\widehat{\nabla l_{H}(\btt)}}$ & -3.78      & -3.46      & -3.51      & -3.60       \\
      \hline
      $\eps_{\widehat{H l_{H} (\btt})}$     & -4.22      & -3.89      & -3.71      & -3.79       \\
      \hline
    \end{tabular}
  \end{center}
\end{table}

\begin{table}[!ht]
  \begin{center}
    \caption{Average precisions (on $\log_{10}$ scale) of the stochastic gradient, expected Fisher
    matrix, and Hessian at $\mle$. \revs{Here  $\eta_g$ and $\eta_{\mathcal{I}}$ are given by
   (\ref{grad_prec_metric}) and (\ref{hess_prec_metric}) respectively.}  \label{tab:stch_prec_mle}}
    \begin{tabular}{|c|c|c|c|c|}
      \hline
                                             & $n=2^{10}$ & $n=2^{11}$ & $n=2^{12}$ & $n=2^{13}$  \\
      \hline
      $\eps_{\widehat{\nabla l_{H}(\btt)}}$  & -0.62  & -1.42  & -0.98  & -1.75     \\
      \hline
      $\eps_{\widehat{H l_{H} (\btt})}$      & -2.37  & -1.86  & -2.13  & -2.04     \\
      \hline
      $\eta_{g}$                             & -0.92  & -0.93  & -0.73  & -0.96     \\
      \hline
      $\eta_{\mathcal{I}}$                   & -1.77  & -1.82  & -1.86  & -2.21     \\
      \hline
    \end{tabular}
  \end{center}
\end{table}

For the relative precision, the interpretation is clear: the estimated gradient and Hessian at
$\btt=(2,2)$ have relative error less than $0.03\%$, which we believe demonstrates their suitability
for numerical optimization. Near the MLE,  the gradient estimate in particular can become less
accurate, meaning that stopping conditions in minimization like $|| \widehat{ \nabla l_{H} ( \btt) }
|| < \eps_{tol}$ may not be suitable, for example, and that if the gradient is sufficiently small,
even the signs of the terms in the estimate may be incorrect, which can be confounding for numerical
optimization.  Nonetheless, in later sections we demonstrate the ability to optimize to the relative
tolerance of $10^{-8}$ in the objective function.

\subsection{Simulated data verifications}
Using the alternative parameterization of the Mat\'ern covariance function with fixed  $\nu = 1$,
which corresponds to a process that is just barely not mean-square differentiable, we simulate five
datasets of size $n=2^{18}$ on an even grid across the domain $[0, 100]^2$ using the $R$ software
\emph{RandomFields} of \cite{Schlather2015}; and we then fit successively larger randomly subsampled
datasets (obtaining both point estimates and approximate $95\%$ confidence intervals via the
expected Fisher information matrix) from each of these of sizes $2^{k}$ for $k$ ranging from $11$ to
$17$, thus working exclusively with irregularly sampled data. We do this for two range parameters of
$\theta_{1} = 5$ and $\theta_{1} = 50$ to further demonstrate the method's flexibility, optimizing
in the weak correlation case with a simple second-order trust-region method that exactly solves the
subproblem as described in \cite{Nocedal99} and with the first-order \emph{method of moving
asymptotes} provided by the NLopt library (\citeauthor{JohnsonNLopt}) in the strongly correlated
case, \revs{as the second-order methods involving the Hessian did not accelerate convergence to the
MLE for the strongly correlated data, which has generally been our experience}.  In both
circumstances, the stopping condition is chosen to be a relative tolerance of $10^{-8}$. For $k \leq
13$, we provide parameters fitted with the exact likelihood to the same tolerance for comparison.
Figures \ref{figure:range_small} and \ref{figure:range_big} summarize the results.  \revs{In
Appendix B, we also provide confidence ellipses from the exact and stochastic expected Fisher
matrices, providing another tool for comparing the inferential conclusions one might reach from the
exact and approximated methods.}

\begin{figure}[!ht] 
  \begin{center}
    \includegraphics[width=\textwidth]{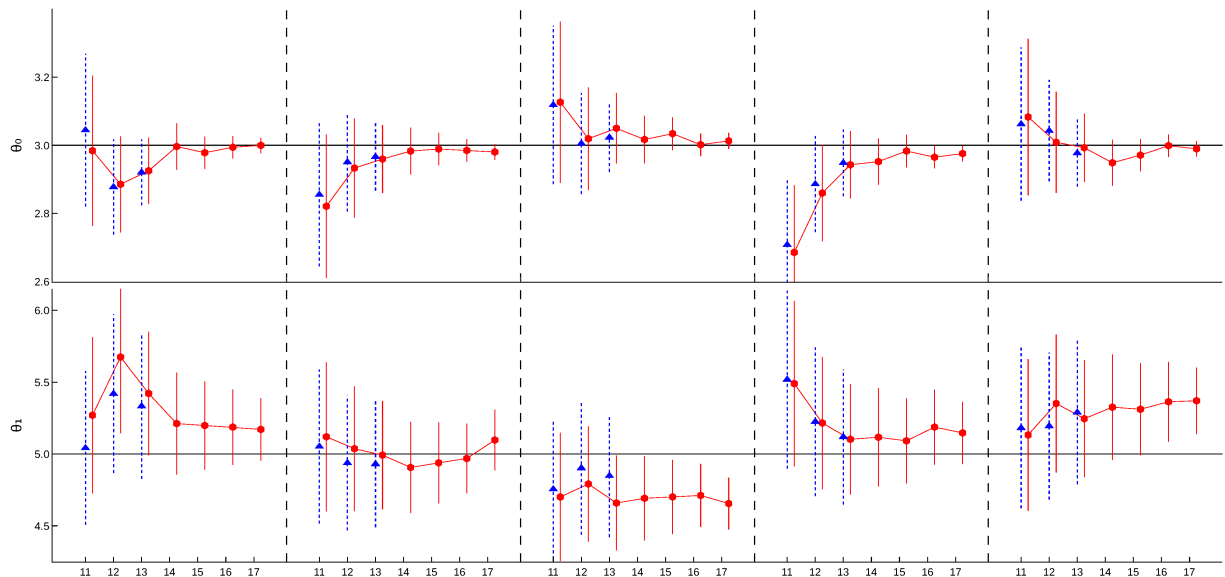}
    \caption{Estimated MLEs and confidence intervals for random subsamplings of $5$ exactly
      simulated datasets of size $n=2^{18}$ with covariance given by (\ref{smatern}) and
      parameters $\btt = (3,5)$ and $\nu=1$, with subsampled sizes $2^{k}$ with $k$ ranging from
      $11$ to $17$ (horizontal axis). Exact estimates provided for the first three sizes with
    triangles. \label{figure:range_small} }
  \end{center}
\end{figure}

\begin{figure}[!ht] 
  \begin{center}
    \includegraphics[width=\textwidth]{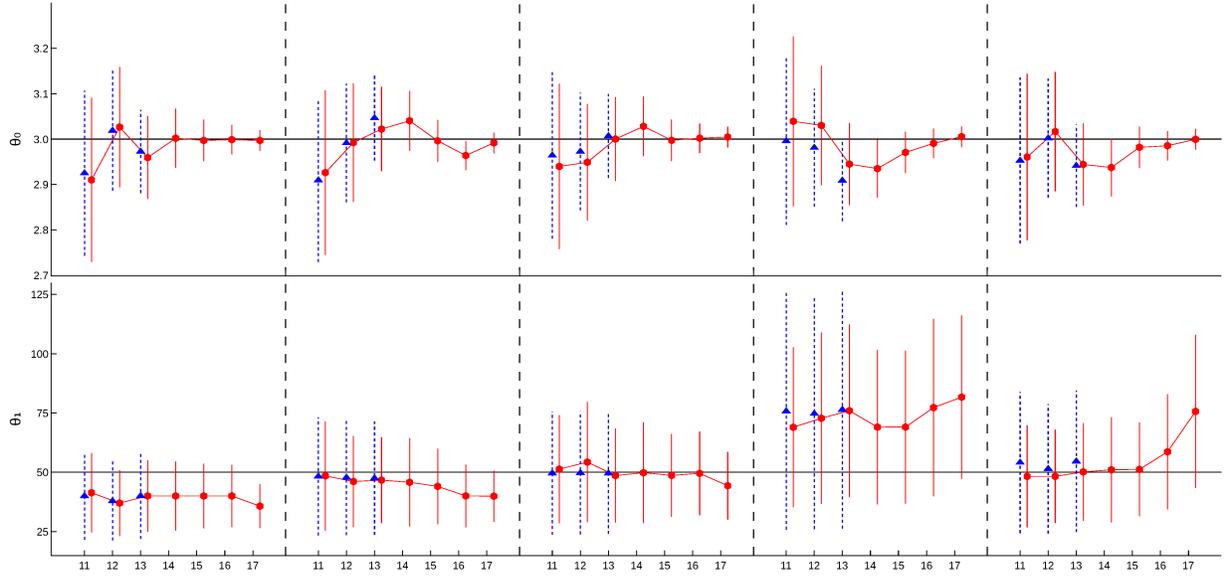}
    \caption{Same as Figure \ref{figure:range_small} except now with a large range
             parameter of $\theta_{1} = 50$.  \label{figure:range_big} }
  \end{center}
\end{figure}

The main conclusion from these simulations is that the minimizers of $-l_{H}$ closely resemble those
of the exact log-likelihood and that the widths of the confidence intervals based on the approximate
method are fairly close to those for the exact method.  Assuming that this degree of accuracy
applies to the larger sample sizes for which we were unable to do exact calculations, we see some
interesting features in the behavior of the estimates as we take larger subsamples of the initial
simulations of $2^{18}$ gridded observations.  Specifically, we see that the estimates of the scale
$\theta_0$ continue to improve as the sample size increases, whereas, especially when the true range
is 50, the estimated ranges do not obviously improve at the larger sample sizes, as is correctly
represented in the almost constant widths of the confidence intervals for these larger sample sizes.
These results are in accord with what is known about fixed-domain asymptotics for the Mat\'ern model
\citep{Stein1999, Zhang2004, Zhang2005}.  We note that the asymptotic results underlying the use of
Fisher information for approximate confidence intervals for MLEs, which includes consistency of the
point estimates, are not valid in this setting.  Nevertheless, the Fisher information matrix still
gives a meaningful measure of uncertainty in the parameter estimates since it corresponds to the
Godambe information matrix of the score equations \citep{Heyde2008}.  Additionally, these results
demonstrate that the stochastic gradient and Hessian estimates are sufficiently stable even near the
MLE and that numerical optimization to a relative tolerance of $10^{-8}$ can be performed.  \revs{It
is worth noting, though, that stochastic optimization to this level of precision was more
expensive in terms of the number of function calls required to reach convergence (which we found
to be about twice as many), although we see no evidence to indicate that that difference will grow with
$n$. For detailed information about the optimization performed to generate Figures
\ref{figure:range_small} and \ref{figure:range_big}, see the Supplementary Material.}

\section{Discussion}

In this paper, we present an approximation of the Gaussian log-likelihood that can be computed in
quasilinear time and  admits both a gradient and Hessian with the same complexity. In the
Numerical results section, we demonstrate that with the exact derivatives of our approximated
covariance matrices and the symmetrized trace estimation we can obtain very stable and performant
estimators for the gradient and Hessian of the approximated log-likelihood and the expected
information matrix. Further, we demonstrate that the minimizers of our approximated likelihood are
nearly identical to the minimizers of the exact likelihood for a wide variety of method parameters
and that the expected information matrix and the Hessian of our approximated likelihood are
relatively close to their exact values. Putting these together, we present a coherent model for computing
approximations of MLEs for Gaussian processes in quasilinear time that is kernel-independent and
avoids many problem-specific computing challenges (such as choosing preconditioners). Further, the
approach we advocate here (and the corresponding software) is flexible, making it an attractive
and fast way to do exploratory work.  

In some circumstances, however, our method is potentially less useful\revs{, as utilizing the
differentiability of the approximation of $\hbvS$} requires first partial derivatives of the
kernel function with respect to the parameters. For simpler models as have been discussed here, this
requirement is not problematic. For some covariance functions, however, these derivatives may need
to be computed with the aid of automatic differentiation (AD) or even finite differencing (FD).
This issue comes up even with the Mat\'ern kernel, since, to the best of our knowledge, no usable
code exists for efficiently computing the derivatives $\frac{\partial^{k}}{\partial \nu^{k}}
\mathcal{K}_{\nu}(x)$ analytically, even though series expansions for these derivatives are
available (see 10.38.2 and 10.40.8 in \cite{NIST}).  Empirically, we have obtained reasonable
estimates for $\hat{\nu}$ using finite difference approximations for these derivatives when $\nu$ is
small. The main difficulty with finite differencing, however, is that it introduces a source of
error that is hard to monitor, so that it can be difficult to recognize when estimates are being
materially affected by the quality of the finite difference approximations. While the algorithm will
still scale with the same complexity if AD or FD is used as a substitute for exactly computed
derivatives, doing so will likely introduce a serious fixed overhead cost as well as potential
numerical error in the latter case.  With that said, however, the overhead will be incurred during
the assembly stage, which is particularly well-suited to extreme parallelization that may mitigate
such performance concerns. 

Another circumstance in which this method may not be the most suitable is one where very accurate
trace estimation is required, such as optimizing to a very high precision. As has been discussed,
the peeling method of \cite{Lin2011} may be used, but matvec actions with the derivative matrices,
especially $\hbvS_{jk}$, have a very high overhead, which may make the peeling method unacceptably
expensive.  Parallelization would certainly also mitigate this cost, but the fact remains that
performing $O(\log n)$ many matvecs with $\hbvS_{jk}$ will come at a significant price.

\revs{In some circumstances, the framework of hierarchical matrices may not be the most appropriate
scientific choice}. It has been shown that, at least in some cases, as the dimension of a problem
increases or its geometry changes, the numerical rank of the low-rank blocks of kernel matrices will
increase \citep{Ambikasaran2014a}, which will affect the scaling of the algorithms that attempt to
control pointwise precision.  For algorithms that do not attempt to control pointwise precision,
such as the one presented here, the complexity will not change, but the quality of the approximation
will deteriorate.  Moreover, off-diagonal blocks of kernel matrices often have low numerical rank
because the corresponding kernel is smooth away from the origin \citep{Ambikasaran2016}.  For
covariance kernels for which this does not hold, off-diagonal blocks may not be of low numerical
rank regardless of the dimension or geometry of the problem. For most standard covariance functions
in spatial and space-time statistics, there is analyticity away from the origin. And most space-time
processes happen in a dimension of at most four, so the problems of dimensionality may not often be
encountered. But for some applications, for example in machine learning, that are done in higher
dimensions, we suggest using care to be sure that the theoretical motivations for this approximation
hold to a reasonable degree. 

\revs{Finally, we note that many of the choices made in this method are subjective and can
potentially be improved. We chose the HODLR format to approximate $\bvS$ due to its simplicity and
transparency with regard to complexity, but there are many other options for matrix compression;
see \cite{Minden2016}, \cite{Chen2017}, and \cite{Wang2018} for three recent and very different
examples of matrix compression. Further, the Nystr\"om approximation for off-diagonal blocks was
chosen so that the covariance matrix would be differentiable, but there are many other methods for
low-rank approximation, and many of them have better theoretical properties; see
\cite{Bebendorf2000}, \cite{Liberty2007} and \cite{Wang2016} for diverse examples of adaptive
low-rank approximations.  These particular methods are not generally differentiable with respect
to kernel parameters \citep{Griewank2008evaluating}, but if one were to prioritize pointwise
approximation quality over differentiability of the approximated covariance matrix, adaptive
low-rank approximation methods like those mentioned above may be reasonable choices.}

\spacingset{1.0}
\section*{Acknowledgments}

This material was based upon work supported by the U.S. Department of Energy, Office of Science,
Office of Advanced Scientific Computing Research (ASCR) under Contracts DE-AC02-06CH11347 and
DE-AC02-06CH11357. We acknowledge partial NSF funding through awards FP061151-01-PR and CNS-1545046
to MA. 

\bibliography{references}

\vspace{-0.15cm}
\begin{flushright}
  \scriptsize \framebox{\parbox{2.5in}{Government License: The submitted manuscript has been created
    by UChicago Argonne, LLC, Operator of Argonne National Laboratory (``Argonne").  Argonne, a U.S.
    Department of Energy Office of Science laboratory, is operated under Contract No.
    DE-AC02-06CH11357.  The U.S. Government retains for itself, and others acting on its behalf, a
    paid-up nonexclusive, irrevocable worldwide license in said article to reproduce, prepare
    derivative works, distribute copies to the public, and perform publicly and display publicly, by or
    on behalf of the Government. The Department of Energy will provide public access to these results of
    federally sponsored research in accordance with the DOE Public Access Plan.
    http://energy.gov/downloads/doe-public-access-plan. }}
	\normalsize
\end{flushright}

\clearpage 

\appendix

\section{Exact expressions for $\hbvS_{jk}$}

As described in Section $2$, derivatives of off-diagonal block approximations in the form of
(\ref{nys}) are given by (\ref{deriv}). The Hessian of the approximated log-likelihood requires the
second derivative of $\hbvS$, which in turn requires the partial derivatives of (\ref{deriv}).
Continuing with the same notation as in Section $2$, three simple product rule computations show
that the $k$th partial derivative of (\ref{deriv}) is given by
\begin{align*} \label{deriv2}
  \hbvS_{j,k,(I,J)} &= \bvS_{j,k,(I,P)} \bvS_{P,P}^{-1} \bvS_{P,J}  \\
  &- \bvS_{j,(I,P)} \bvS_{P,P}^{-1} \bvS_{k, (P,P)} \bvS_{P, P}^{-1} \bvS_{P, J} \\
  &+ \bvS_{j,(I,P)} \bvS_{P,P}^{-1} \bvS_{k, (P,J)} \\
  &+ \bvS_{k,(I,P)} \bvS_{(P,P)}^{-1} \bvS_{j, (P,P)} \bvS_{P, P}^{-1} \bvS_{P, J} \\
  &- \bvS_{I,P} \bvS_{(P,P)}^{-1} \bvS_{k, (P,P)} \bvS_{(P,P)}^{-1} \bvS_{j, (P,P)} \bvS_{P, P}^{-1} \bvS_{P, J} \\
  &+ \bvS_{(I,P)} \bvS_{(P,P)}^{-1} \bvS_{j,k, (P,P)} \bvS_{P, P}^{-1} \bvS_{P, J} \\
  &- \bvS_{I,P} \bvS_{(P,P)}^{-1} \bvS_{j, (P,P)} \bvS_{(P,P)}^{-1} \bvS_{k, (P,P)} \bvS_{P, P}^{-1} \bvS_{P, J} \\
  &+ \bvS_{(I,P)} \bvS_{(P,P)}^{-1} \bvS_{j,(P,P)} \bvS_{P, P}^{-1} \bvS_{k, (P, J)} \\
  &+ \bvS_{k,(I,P)} \bvS_{P,P}^{-1} \bvS_{j, (P,J)}  \\
  &- \bvS_{(I,P)} \bvS_{k, (P,P)}^{-1} \bvS_{k, (P,P)} \bvS_{P, P}^{-1} \bvS_{j, (P, J)} \\
  &+ \bvS_{(I,P)} \bvS_{P,P}^{-1} \bvS_{j,k, (P,J)}.
\end{align*}
Although this expression looks unwieldy and expensive, each line is still expressible as the sum
of rank $p$ matrices that can be written as $\bm{U} \bm{S} \bm{V}^{T}$, where $\bm{S} \in
\mathbb{R}^{p \times p}$, meaning that a matvec operation with the block shown above will still
scale with linear complexity for fixed $p$. While the overhead involved is undeniably substantial,
the assembly and application of $\hbvS_{jk}$ is nonetheless demonstrated to scale with quasilinear
complexity.

\end{document}